# A NOVEL APPROACH FOR SECURE DATA AGGREGATION IN WIRELESS SENSOR NETWORKS

Vivaksha Jariwala[1], Devesh Jinwala[2]

[1] Lecturer, Department of Computer Engineering, C. K. Pithawalla College of Engineering and Technology, Near Malvan Mandir, Via Magdalla Port, Surat- Dumas road, Surat – 395007, Gujarat , India
vivakshajariwala@gmail.com

[2] Associate Professor, Department of Computer Engineering, S. V. National Institute of Technology, Ichchhanath, Surat – 395007, Gujarat, India
dcjinwala@gmail.com

**Abstract:** *The Wireless Sensor Networks (WSNs) are composed of resource starved sensor nodes that are deployed to sense, process and communicate vital information to the base station. Due to the stringent constraints on the resources in the sensor nodes on one hand and due to the communications costs being always significantly higher than the data processing costs, the WSNs typically, employ in-network processing, which aims at reducing effectively, the total number of packets eventually transmitted to the base station. Such in-network processing largely employs data aggregation operations that aggregate the data into a compact representation for further transmission. However, due to the ubiquitous & pervasive deployment, heavier resource demands of the security protocols and due to the stringent resource constraints in WSN nodes, the security concerns in WSNs are even otherwise critical. These concerns assume alarming proportions when using data aggregation in which the output of the data aggregator nodes depends on that of various other nodes. Hence, the protocols for data aggregation have to carefully devised with a constant vigil on ensuring security of the data. In this paper, based on our survey of the existing research efforts for ensuring secure data aggregation, we propose a novel approach using homomorphic encryption and additive digital signatures to achieve confidentiality, integrity and availability for secure data aggregation in wireless sensor networks.*

# 1   INTRODUCTION

Wireless Sensor Networks (WSNs) consist of hundreds or thousands of tiny sensing devices with restricted memory, computational and communication resources [1]. These devices are severely resource constrained with a typical sensor mote consisting of only 8-bit, 4 MHz processor, 128 kb program flash memory, 512 kb of EEPROM and 2 X AA batteries [2].

The potential applications of the WSNs typically range from those in defense, military, environmental monitoring, health monitoring, home appliances, civilian societal surveillance applications etc. [1]. However, primarily due to the severe constraints and secondarily due to the inherent increased costs in communications relative to that in processing, the WSNs follow in-network processing, wherein the emphasis is on *on-the-fly* processing of the data packets before being communicated eventually to the base station [3]. The processing typically consists of the data aggregation operations like summarization, summation, averaging, finding the minimum/maximum of a set of sensed values. The eventual advantage of data aggregation is the overall reduction on the total number of packets transmitted to the base station resulting in conservation of energy and bandwidth.

However, while being advantaged in a manner described above, the data aggregation operation indeed gives rise to other consequences. Data aggregator nodes usually collect data from the sensor nodes, apply aggregation operations on it and eventually communicate the processed data to base station. Thus, the sanctity of the data values communicated eventually to the base station lies on (1) the sanctity of the data values passed on the aggregator node (2) the sanctity of the aggregation operations done at the aggregator node. Obviously, in presence of the malicious nodes, neither the aggregated data nor the aggregation operation remains trustworthy. Thus, it is essential to ensure that the data aggregation operational paradigm is inherently secure.

In general, when designing a secure data aggregation protocol, the primary objective to devise a secure aggregation function that computes the data aggregates securely and the secondary objective is to ensure that other than the sink and the sources, no intermediate node should have any knowledge of the raw data or the aggregation result.

One can indeed find attempts in the literature to devise the secure data aggregation protocols as we survey further in section 2. However, with the motivation to improve upon the same, in this paper, we propose a novel approach for secure data aggregation that provides confidentiality and integrity of aggregated data, at an improved performance.

The rest of the paper is organized thus: in section 2, we describe the essential background and the related attempts in ensuring in secure data aggregation in wireless sensor networks. We describe the proposed approach for secure data aggregation in section 3 and analyze the same in section 4, showing the benefits gained, whereas we conclude with the scope of future work in section 5.

## 2   THEORETICAL BACKGROUND AND RELATED WORK

### 2.1   Data Aggregation

Wireless Sensor Network consists of a number of sensor nodes that sense the data from environment process it and communicate the processed data to the base station. In WSNs, sensor nodes are really energy constrained. Thus, as mentioned earlier, if all the sensor node transmit their information to the base station then precious energy of WSNs is wasted and life time of WSNs is decreased. Moreover, in WSNs, neighboring sensor nodes usually sense same or co-related data which is semantically redundant and hence does not need all to be communicated to the base station. Hence, instead of communicating the redundant data, a better alternative is to identify and pre-process such data into an aggregated or summarized form and communicate it. Such on-the-fly pre-processing (in-network processing) of the sensed data eventually helps reducing the total number of packets transmitted to the base station, lowering the communications cost and thereby reducing the energy overhead. The typical pre-processing operations (data aggregation) are finding the maximum, minimum, averaging, duplicate eliminating etc [4]. However, as mentioned before, just because data aggregation operations necessitate the intermediate nodes actually 'see' the data values, it gives rise to the security and privacy issues thereby necessitating the Secure Data Aggregation operations.

## 2.2  Secure Data Aggregation

As, almost all the applications of WSNs demand certain level of security it is mandatory not to sacrifice security for data aggregation operations. Normally, one way of achieving security is to impose confidentiality of the data, that in turn is achieved by encrypting it. However, this also means that each intermediate node (that acts as a data aggregator) needs to decrypt the incoming data before applying any aggregation operation on it. Such multiple rounds of decryption-encryption at each intermediate node is obviously more vulnerable to attacks as compared to the *end-to-end* notion of security. Hence, it is necessary to explore carefully the probable approaches for incorporating security in the WSNs based on the data aggregation operational paradigm [5][6].

Of late, with the research results revealing the possibility of the *ciphertext processing* (as compared to the default recourse to *plaintext processing*), one of the ways to make secure data aggregates is to use such ciphertext processing protocol in WSNs. Such ciphertext processing based approaches are known as those based on privacy homomorphism [7]. The fundamental basis for data aggregations are cryptographic methods that provide privacy homomorphism property. A privacy homomorphism (PH) is an encryption transformation that allows direct computation on encrypted data [7].

Let Q and R denote two rings, and + and $\oplus$ denote addition operations on the rings. Let k denotes the key space. We denote an encryption transformation E : K X Q →R and the corresponding decryption transformation D : K X R → Q. Given a, b $\epsilon$ Q and k, k1, k2 $\epsilon$ K, we term a + b = Dk (Ek(a) $\oplus$ Ek(b)) additively homomorphic with a single secret key and a + b = Dk (k1,k2) (Ek1(a) $\oplus$ Ek2(b)) additively homomorphic with multiple secret keys. We denote an asymmetric additively homomorphic encryption transformation as a + b = Dp (Ep(a) ⊕ Eq(b)) with (p,q) being a private, public key pair.

In Fig. 1, we show the probable alternatives for ensuring secure data aggregation in the wireless sensor networks.

Thus, secure data aggregation can be achieved in two ways viz. with plain sensor data and with encrypted sensor data. Our focus here is on the latter approach – the one employing privacy homomorphism. However, the latter approach too, further can be classified into two sub approaches viz.  one based on Symmetric Key Cryptography (SKC) OR the one employing  Asymmetric Key Cryptography (AKC/PKC).

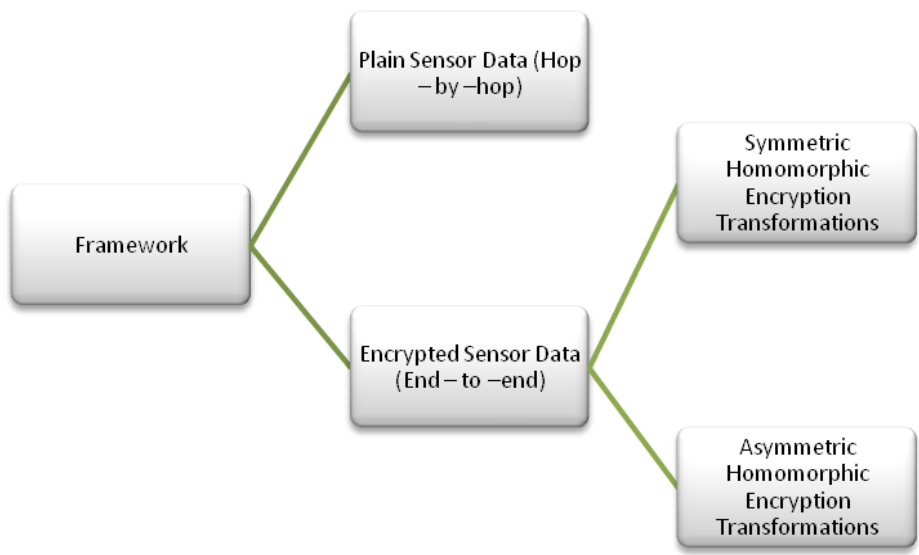

**Fig. 1.** Framework for Secure Data Aggregation in WSNs

## 2.3   Related Work

Based on the survey of the contemporary work on privacy homomorphism, we observe that there are very few schemes that employ the PKC. PKC algorithms anyway have been known to be too resource intensive and hence infeasible to be used directly on the resource constrained WSNs. Moreover, the expansion in bit size during the transformation of plain text to cipher text introduces costly communication overhead.

Hence it is essential to investigate alternatives that feasibly implement the PKC algorithms. In [8], the authors investigate the feasibility of employing various suitable public-key schemes including using elliptic curve operations on the Mica2 sensor nodes [2]. We summarize the all such approaches for PKC in Table 1 below.

In this research exercise, we attempt to explore further the feasibility of the PKC schemes using privacy homomorphism in WSNs. As we explain further in the next section, we employ digital signatures to achieve secure data aggregation, providing confidentiality and integrity of data. As all the approach proposed till now for secure data aggregation, many few of them are using PKC for achieving secure data aggregation as shown in the table below. Our novel approach combines the idea from the PKC based Discrete Logarithms approach OU proposed in [9] and ECDSA proposed in [9] to achieve confidentiality and integrity of data to be aggregated and passed to the base station. So our proposed algorithm would combine preeminent features of PKC based OU and ECDSA to give more efficient aggregate result. To the best of our knowledge that is not done till now and hence it is would be worth exploring it.

**Table 1.** Asymmetric homomorphic encryption transformations

| Approaches | Citation | Remark |
|---|---|---|
| Discrete Logarithms | [10] | OU- As secure as factoring and based on the ability of computing discrete logarithms in particular subgroup. |
| Probabilistic Cryptosystem | [11] | Benaloh - Encryption cost is dependent on the size of the plaintext. Additive homomorphic property is achieved through the multiplication of cipher texts. |

| Elliptic Curve Cryptography | [12] | EC-NS - factoring based algorithms are exported to particular families of EC. EC-P - is not as efficient and requires too much computation. EC-EG - additively homomorphic, and ciphertexts are combined through addition. |
| Group Based | [13] | HCDA - sensor nodes in a group use the same public key. It is based on Elliptic curve cryptography so it is not affected by node compromise attacks whereas symmetric key based protocols are significantly affected from these attacks. |
| Hierarchical | [14] | Confidentiality and integrity is provided. |

## 3 PROPOSED APPROACH

As mentioned above, we use Elliptic Curve Okamoto Uchiyama (EC-OU) [10] scheme for confidentiality. Homomorphic encryption does not provide integrity. Therefore, we use public key elliptic curve cryptography and digital signatures (ECDSA)[9] to provide integrity. In our scheme, we have proposed two algorithms, first to be implemented on sensor node and the other one to be implemented on the base station.

Our proposed algorithm for sensor node generates unique signature for each outgoing message say **Sig(x)**. The data from each sensor node is encrypted using the base station's public key viz. **Enc(x)**. Thus, every node communicates the encrypted message, its signature and its public key to its parent. After receiving the message from every child, the parent combines signature, public key and encrypted message and sums it. The message so received can be decrypted only by the base station that has the corresponding private key. Thus, the base station can now verify the sum of the signatures given the sum of the public keys. The psuedocode for both the proposed algorithms are shown below:

```
Algorithm 1 : SensorNodeAlgo(mi, zi, Q, k)
// Maps its reading mi on the elliptic curve D
// Elliptic Curve Parameters D = (q, FR, a, b, T, p, h)
// Each sensor node will computes following
1. zi * T = (x,y), it is public key
2. R =(r(x),r(y)) = k * T
3. V = k-1 mod p
4. Signature si= k-1 (mi +zi * r(x)) mod p
5. Call EC-OU(mi)
6. mi' = enc(mi)
7. if sensor is a parent
   s = ∑ si // senor combines the signature
   m = ∑ mi // combines all cipher texts into one cipher text
end if
```

```
Algorithm 2 : BaseStationAlgo(mi, si, qi,Z, k)
// Maps its reading mi on the elliptic curve D
// Elliptic Curve Parameters D = (q, FR, a, b, T, p, h)
// Each sensor node will computes following
1. call EC-OU (mi')
2. ∑ mi' = Dec(∑mi)
3. r = (r(x),r(y)) = k * T
4. w=s-1 mod p
```

```
5. u₁ = mw mod p and u₂=r(x)w mod p
6. X = u₁T + u₂Z
7. j = X(x) mod p
8. if j ==r then
     • Signature is verified
   end if
```

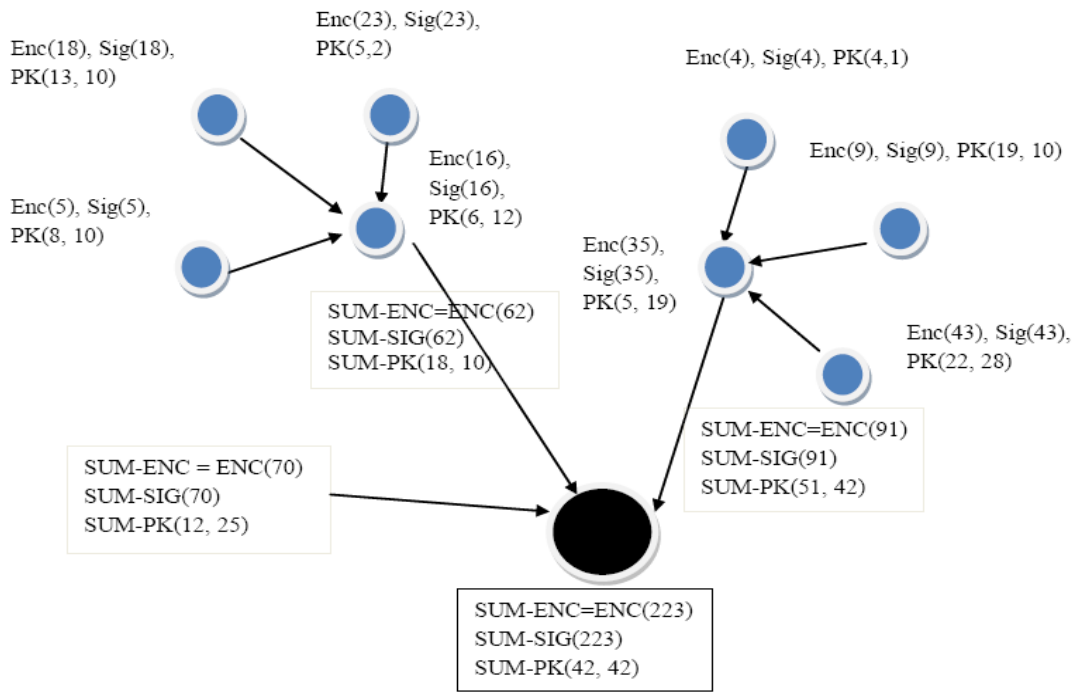

**Fig. 2.** Example

## 4   ANALYSIS

Our proposed approach is based on ECDSA [9]. We have used ECDSA to provide integrity of data for data aggregation. ECDSA is assumed to be secure under the assumption that the underlying group is generic and that a collision resistant hash function has been used.

We have used EC-OU [10] for asymmetric homomorphic encryption transformation that gives us the confidentiality of data. EC-OU is provably secure and has many properties like: (a) Its trapdoor technique is essentially different from any other asymmetric schemes. (b) It is a probabilistic encryption scheme. (c) It can be proven to be as secure as the intractability of factoring $\mathbf{n = p^2q}$ (d) It is semantically secure under the p-subgroup assumption, which is comparable to the quadratic residue and higher degree residue assumptions. (e) It has homomorphic property. Elliptic curve cryptosystem is also having many advantage like (a) their use of small keys which lead to short ciphertexts, (b) the smaller real-estate required for hardware implementations (number of gates) and (c) a better security-per-bit ratio. By considering advantage and efficiency of OU and Elliptic curve cryptosystem, we have chosen EC-OU that can combine advantage of both OU as well as Elliptic curve cryptosystem.

## 5   CONCLUSION AND FUTURE WORK

We propose a novel approach for secure data aggregation in WSNs. The proposed approach uses homomorphic encryption EC-OU algorithm to achieve data confidentiality while allowing in-network aggregation. We have used an additively digital signature algorithm based on Elliptic Curve Digital Signature Algorithm (ECDSA) to achieve integrity of the aggregate. To the best of our knowledge, combining preeminent feature of EC-OU and ECDSA to attain confidentiality and integrity and digging up benefit of both is not done till now and hence it would be worth exploring. Our future work will include analysis of our novel approach and implementing the same in TinyOS/TOSSIM .

**Acknowledgments**. The authors would like to thank the anonymous reviewers for their comments which greatly enhanced this paper.